\documentclass[letterpaper,compsoc,twoside]{IEEEtran}
\usepackage{fixltx2e} 
\usepackage{cmap} 
\usepackage{ifthen}
\usepackage[T1]{fontenc}
\usepackage[utf8]{inputenc}
\usepackage{amsmath}

\usepackage[font={small,it},labelfont=bf]{caption}
\usepackage{float}

\pdfoutput=1
\usepackage{scipy}
\makeatletter
\def\PY@reset{\let\PY@it=\relax \let\PY@bf=\relax%
    \let\PY@ul=\relax \let\PY@tc=\relax%
    \let\PY@bc=\relax \let\PY@ff=\relax}
\def\PY@tok#1{\csname PY@tok@#1\endcsname}
\def\PY@toks#1+{\ifx\relax#1\empty\else%
    \PY@tok{#1}\expandafter\PY@toks\fi}
\def\PY@do#1{\PY@bc{\PY@tc{\PY@ul{%
    \PY@it{\PY@bf{\PY@ff{#1}}}}}}}
\def\PY#1#2{\PY@reset\PY@toks#1+\relax+\PY@do{#2}}

\expandafter\def\csname PY@tok@gd\endcsname{\def\PY@tc##1{\textcolor[rgb]{0.63,0.00,0.00}{##1}}}
\expandafter\def\csname PY@tok@gu\endcsname{\let\PY@bf=\textbf\def\PY@tc##1{\textcolor[rgb]{0.50,0.00,0.50}{##1}}}
\expandafter\def\csname PY@tok@gt\endcsname{\def\PY@tc##1{\textcolor[rgb]{0.00,0.27,0.87}{##1}}}
\expandafter\def\csname PY@tok@gs\endcsname{\let\PY@bf=\textbf}
\expandafter\def\csname PY@tok@gr\endcsname{\def\PY@tc##1{\textcolor[rgb]{1.00,0.00,0.00}{##1}}}
\expandafter\def\csname PY@tok@cm\endcsname{\let\PY@it=\textit\def\PY@tc##1{\textcolor[rgb]{0.25,0.50,0.56}{##1}}}
\expandafter\def\csname PY@tok@vg\endcsname{\def\PY@tc##1{\textcolor[rgb]{0.73,0.38,0.84}{##1}}}
\expandafter\def\csname PY@tok@m\endcsname{\def\PY@tc##1{\textcolor[rgb]{0.13,0.50,0.31}{##1}}}
\expandafter\def\csname PY@tok@mh\endcsname{\def\PY@tc##1{\textcolor[rgb]{0.13,0.50,0.31}{##1}}}
\expandafter\def\csname PY@tok@cs\endcsname{\def\PY@tc##1{\textcolor[rgb]{0.25,0.50,0.56}{##1}}\def\PY@bc##1{\setlength{\fboxsep}{0pt}\colorbox[rgb]{1.00,0.94,0.94}{\strut ##1}}}
\expandafter\def\csname PY@tok@ge\endcsname{\let\PY@it=\textit}
\expandafter\def\csname PY@tok@vc\endcsname{\def\PY@tc##1{\textcolor[rgb]{0.73,0.38,0.84}{##1}}}
\expandafter\def\csname PY@tok@il\endcsname{\def\PY@tc##1{\textcolor[rgb]{0.13,0.50,0.31}{##1}}}
\expandafter\def\csname PY@tok@go\endcsname{\def\PY@tc##1{\textcolor[rgb]{0.20,0.20,0.20}{##1}}}
\expandafter\def\csname PY@tok@cp\endcsname{\def\PY@tc##1{\textcolor[rgb]{0.00,0.44,0.13}{##1}}}
\expandafter\def\csname PY@tok@gi\endcsname{\def\PY@tc##1{\textcolor[rgb]{0.00,0.63,0.00}{##1}}}
\expandafter\def\csname PY@tok@gh\endcsname{\let\PY@bf=\textbf\def\PY@tc##1{\textcolor[rgb]{0.00,0.00,0.50}{##1}}}
\expandafter\def\csname PY@tok@ni\endcsname{\let\PY@bf=\textbf\def\PY@tc##1{\textcolor[rgb]{0.84,0.33,0.22}{##1}}}
\expandafter\def\csname PY@tok@nl\endcsname{\let\PY@bf=\textbf\def\PY@tc##1{\textcolor[rgb]{0.00,0.13,0.44}{##1}}}
\expandafter\def\csname PY@tok@nn\endcsname{\let\PY@bf=\textbf\def\PY@tc##1{\textcolor[rgb]{0.05,0.52,0.71}{##1}}}
\expandafter\def\csname PY@tok@no\endcsname{\def\PY@tc##1{\textcolor[rgb]{0.38,0.68,0.84}{##1}}}
\expandafter\def\csname PY@tok@na\endcsname{\def\PY@tc##1{\textcolor[rgb]{0.25,0.44,0.63}{##1}}}
\expandafter\def\csname PY@tok@nb\endcsname{\def\PY@tc##1{\textcolor[rgb]{0.00,0.44,0.13}{##1}}}
\expandafter\def\csname PY@tok@nc\endcsname{\let\PY@bf=\textbf\def\PY@tc##1{\textcolor[rgb]{0.05,0.52,0.71}{##1}}}
\expandafter\def\csname PY@tok@nd\endcsname{\let\PY@bf=\textbf\def\PY@tc##1{\textcolor[rgb]{0.33,0.33,0.33}{##1}}}
\expandafter\def\csname PY@tok@ne\endcsname{\def\PY@tc##1{\textcolor[rgb]{0.00,0.44,0.13}{##1}}}
\expandafter\def\csname PY@tok@nf\endcsname{\def\PY@tc##1{\textcolor[rgb]{0.02,0.16,0.49}{##1}}}
\expandafter\def\csname PY@tok@si\endcsname{\let\PY@it=\textit\def\PY@tc##1{\textcolor[rgb]{0.44,0.63,0.82}{##1}}}
\expandafter\def\csname PY@tok@s2\endcsname{\def\PY@tc##1{\textcolor[rgb]{0.25,0.44,0.63}{##1}}}
\expandafter\def\csname PY@tok@vi\endcsname{\def\PY@tc##1{\textcolor[rgb]{0.73,0.38,0.84}{##1}}}
\expandafter\def\csname PY@tok@nt\endcsname{\let\PY@bf=\textbf\def\PY@tc##1{\textcolor[rgb]{0.02,0.16,0.45}{##1}}}
\expandafter\def\csname PY@tok@nv\endcsname{\def\PY@tc##1{\textcolor[rgb]{0.73,0.38,0.84}{##1}}}
\expandafter\def\csname PY@tok@s1\endcsname{\def\PY@tc##1{\textcolor[rgb]{0.25,0.44,0.63}{##1}}}
\expandafter\def\csname PY@tok@gp\endcsname{\let\PY@bf=\textbf\def\PY@tc##1{\textcolor[rgb]{0.78,0.36,0.04}{##1}}}
\expandafter\def\csname PY@tok@sh\endcsname{\def\PY@tc##1{\textcolor[rgb]{0.25,0.44,0.63}{##1}}}
\expandafter\def\csname PY@tok@ow\endcsname{\let\PY@bf=\textbf\def\PY@tc##1{\textcolor[rgb]{0.00,0.44,0.13}{##1}}}
\expandafter\def\csname PY@tok@sx\endcsname{\def\PY@tc##1{\textcolor[rgb]{0.78,0.36,0.04}{##1}}}
\expandafter\def\csname PY@tok@bp\endcsname{\def\PY@tc##1{\textcolor[rgb]{0.00,0.44,0.13}{##1}}}
\expandafter\def\csname PY@tok@c1\endcsname{\let\PY@it=\textit\def\PY@tc##1{\textcolor[rgb]{0.25,0.50,0.56}{##1}}}
\expandafter\def\csname PY@tok@kc\endcsname{\let\PY@bf=\textbf\def\PY@tc##1{\textcolor[rgb]{0.00,0.44,0.13}{##1}}}
\expandafter\def\csname PY@tok@c\endcsname{\let\PY@it=\textit\def\PY@tc##1{\textcolor[rgb]{0.25,0.50,0.56}{##1}}}
\expandafter\def\csname PY@tok@mf\endcsname{\def\PY@tc##1{\textcolor[rgb]{0.13,0.50,0.31}{##1}}}
\expandafter\def\csname PY@tok@err\endcsname{\def\PY@bc##1{\setlength{\fboxsep}{0pt}\fcolorbox[rgb]{1.00,0.00,0.00}{1,1,1}{\strut ##1}}}
\expandafter\def\csname PY@tok@kd\endcsname{\let\PY@bf=\textbf\def\PY@tc##1{\textcolor[rgb]{0.00,0.44,0.13}{##1}}}
\expandafter\def\csname PY@tok@ss\endcsname{\def\PY@tc##1{\textcolor[rgb]{0.32,0.47,0.09}{##1}}}
\expandafter\def\csname PY@tok@sr\endcsname{\def\PY@tc##1{\textcolor[rgb]{0.14,0.33,0.53}{##1}}}
\expandafter\def\csname PY@tok@mo\endcsname{\def\PY@tc##1{\textcolor[rgb]{0.13,0.50,0.31}{##1}}}
\expandafter\def\csname PY@tok@mi\endcsname{\def\PY@tc##1{\textcolor[rgb]{0.13,0.50,0.31}{##1}}}
\expandafter\def\csname PY@tok@kn\endcsname{\let\PY@bf=\textbf\def\PY@tc##1{\textcolor[rgb]{0.00,0.44,0.13}{##1}}}
\expandafter\def\csname PY@tok@o\endcsname{\def\PY@tc##1{\textcolor[rgb]{0.40,0.40,0.40}{##1}}}
\expandafter\def\csname PY@tok@kr\endcsname{\let\PY@bf=\textbf\def\PY@tc##1{\textcolor[rgb]{0.00,0.44,0.13}{##1}}}
\expandafter\def\csname PY@tok@s\endcsname{\def\PY@tc##1{\textcolor[rgb]{0.25,0.44,0.63}{##1}}}
\expandafter\def\csname PY@tok@kp\endcsname{\def\PY@tc##1{\textcolor[rgb]{0.00,0.44,0.13}{##1}}}
\expandafter\def\csname PY@tok@w\endcsname{\def\PY@tc##1{\textcolor[rgb]{0.73,0.73,0.73}{##1}}}
\expandafter\def\csname PY@tok@kt\endcsname{\def\PY@tc##1{\textcolor[rgb]{0.56,0.13,0.00}{##1}}}
\expandafter\def\csname PY@tok@sc\endcsname{\def\PY@tc##1{\textcolor[rgb]{0.25,0.44,0.63}{##1}}}
\expandafter\def\csname PY@tok@sb\endcsname{\def\PY@tc##1{\textcolor[rgb]{0.25,0.44,0.63}{##1}}}
\expandafter\def\csname PY@tok@k\endcsname{\let\PY@bf=\textbf\def\PY@tc##1{\textcolor[rgb]{0.00,0.44,0.13}{##1}}}
\expandafter\def\csname PY@tok@se\endcsname{\let\PY@bf=\textbf\def\PY@tc##1{\textcolor[rgb]{0.25,0.44,0.63}{##1}}}
\expandafter\def\csname PY@tok@sd\endcsname{\let\PY@it=\textit\def\PY@tc##1{\textcolor[rgb]{0.25,0.44,0.63}{##1}}}

\def\PYZus{\char`\_}

\def\PYZgt{\char`\>}

\def\PYZhy{\char`\-}

\def\PYZdq{\char`\"}

\makeatother

\providecommand*{\DUrole}[2]{%
  \ifcsname DUrole#1\endcsname%
    \csname DUrole#1\endcsname{#2}%
  \else
    \ifcsname docutilsrole#1\endcsname%
      \csname docutilsrole#1\endcsname{#2}%
    \else%
      #2%
    \fi%
  \fi%
}

\ifthenelse{\isundefined{\hypersetup}}{
  \usepackage[colorlinks=true,linkcolor=blue,urlcolor=blue]{hyperref}
  \urlstyle{same} 
}{}

\begin{document}
\newcounter{footnotecounter}\title{modernizing PHCpack through phcpy}\author{Jan Verschelde$^{\setcounter{footnotecounter}{1}\fnsymbol{footnotecounter}\setcounter{footnotecounter}{2}\fnsymbol{footnotecounter}}$%
          \setcounter{footnotecounter}{1}\thanks{\fnsymbol{footnotecounter} %
          Corresponding author: \protect\href{mailto:jan@math.uic.edu}{jan@math.uic.edu}}\setcounter{footnotecounter}{2}\thanks{\fnsymbol{footnotecounter} University of Illinois at Chicago}\thanks{%

          \noindent%
          Copyright\,\copyright\,2014 Jan Verschelde. This is an open-access article distributed under the terms of the Creative Commons Attribution License, which permits unrestricted use, distribution, and reproduction in any medium, provided the original author and source are credited. http://creativecommons.org/licenses/by/3.0/%
        }}\maketitle
          \renewcommand{\leftmark}{PROC. OF THE 6th EUR. CONF. ON PYTHON IN SCIENCE (EUROSCIPY 2013)}
          \renewcommand{\rightmark}{MODERNIZING PHCPACK THROUGH PHCPY}

\setcounter{page}{71}
\newcommand*{\docutilsroleref}{\ref}
\newcommand*{\docutilsrolelabel}{\label}
\AtEndDocument{\cleardoublepage}
\begin{abstract}PHCpack is a large software package for solving systems of polynomial
equations. The executable phc~is menu driven and file oriented. This
paper describes the development of phcpy, a Python interface to PHCpack.
Instead of navigating through menus, users of phcpy~solve systems in the
Python shell or via scripts. Persistent objects replace intermediate
files.\end{abstract}

\section{Introduction%
  \label{introduction}%
}

Our mathematical problem is to solve a system of polynomial equations in
several variables. The discrete part of the output data includes the
number of solutions and degrees of positive dimensional solution sets.
When the input is exact or if the coefficients of the polynomial can be
given with any accuracy, the isolated solutions can be then approximated
up to any accuracy and for each positive dimensional solution component,
as many generic points as its degree can be computed.

Version 1.0 of PHCpack was archived in~\cite{Ver99}. PHCpack incorporates two
external software packages: MixedVol~\cite{GLW05} and QDlib~\cite{HLB01}.
Although the original focus was to approximate all isolated
complex solutions, PHCpack
prototyped many of the early algorithms in numerical algebraic
geometry~\cite{SVW03}, \cite{SVW05}. Recent updates are listed in~\cite{Ver10}.

The Python interface to PHCpack got a first start when Kathy Piret met
William Stein at the software for algebraic geometry workshop at the
Institute for Mathematics and its Applications in Minneapolis in the
Fall of 2006. The first version of this interface is described in \cite{Pir08}.
Sage \cite{S+13}~ offers the interface phc.py,
developed by William Stein, Marshall Hampton and Alex Jokela.
Version 0.0.1 of phcpy~originated at lecture 40 of the author
in the graduate course MCS 507 in the Fall of 2012,
as an illustration of Sphinx~\cite{Bra}.
Version 0.1.0 was prepared for presentation at EuroSciPy 2013 (August 2013),
version 0.1.4 corresponds to the first version of this paper.
The current version of phcpy~is~0.1.5.

We first outline in the next section the application of numerical
homotopy continuation methods to compute, over the complex numbers,
all isolated solutions and all
positive dimensional irreducible solution sets of a polynomial system.
Then we describe how phcpy~relates to other interfaces to PHCpack. The
functionality of phcpy~is then summarized briefly as the online Sphinx
documentation is more extensive and still growing. The Python interface
to PHCpack builds directly on the C interface to the Ada code.

Related software packages that apply homotopy continuation methods to
solve polynomial systems are (in alphabetical order):
Bertini~\cite{BHSW}, \cite{BHSW08},
HOMPACK90~\cite{WSM+97} (the successor of HOMPACK \cite{WBM87}),
HOM4PS \cite{GLL02}, HOM4PS-2.0~\cite{LLT08}, NAG4M2~\cite{Ley11},
PHoM~\cite{GKK+04}, and pss3.0.5~\cite{Mal}.
As polynomial homotopy continuation methods
involve many algorithms from various fields of computing, every software
has its unique strengths and the statement “\emph{no one package provides all
such capabilities}” quoted from~\cite{BHSW08} remains true today.

\section{Polynomial Homotopy Continuation%
  \label{polynomial-homotopy-continuation}%
}

Our mathematical problem is to solve a polynomial
system~$f({\bf x}) = {\bf 0}$ in several
variables~${\bf x}= (x_1,x_2,\ldots,x_n)$. A homotopy connects
$f({\bf x}) = {\bf 0}$ to a system $g({\bf x}) = {\bf 0}$
with known solutions:\begin{equation*}
h({\bf x},t) = \gamma (1-t) g({\bf x}) + t f({\bf x}) = {\bf 0},
\quad \gamma \in {\mathbb C}.
\end{equation*}For almost all values for $\gamma$, the solutions of
$h({\bf x},t) = {\bf 0}$ are regular for all $t \in [0,1)$.
Numerical continuation methods track solution paths defined
by~$h({\bf x},t) = {\bf 0}$.

For systems with natural parameters $\lambda$, we
solve $f({\mbox{\boldmath $\lambda$}},{\bf x}) = {\bf 0}$ first
for generic values of the parameters
${\mbox{\boldmath $\lambda$}}= {\mbox{\boldmath $\lambda$}}_0$ and
then use\begin{equation*}
h({\bf x},t) = (1-t) f({\mbox{\boldmath $\lambda$}}_0,{\bf x})
+ t f({\mbox{\boldmath $\lambda$}}_1,{\bf x}) = {\bf 0},
\end{equation*}to solve a specific instance
$f({\mbox{\boldmath $\lambda$}}_1,{\bf x}) = {\bf 0}$.

The schematic in Figure~\DUrole{ref}{figcomplexparcon} illustrates that singular
solutions along the paths are avoided by a generic choice of the
parameters~${\mbox{\boldmath $\lambda$}}_0$ at~$t=0$.\begin{figure}[]\noindent\makebox[\columnwidth][c]{\includegraphics[width=\columnwidth]{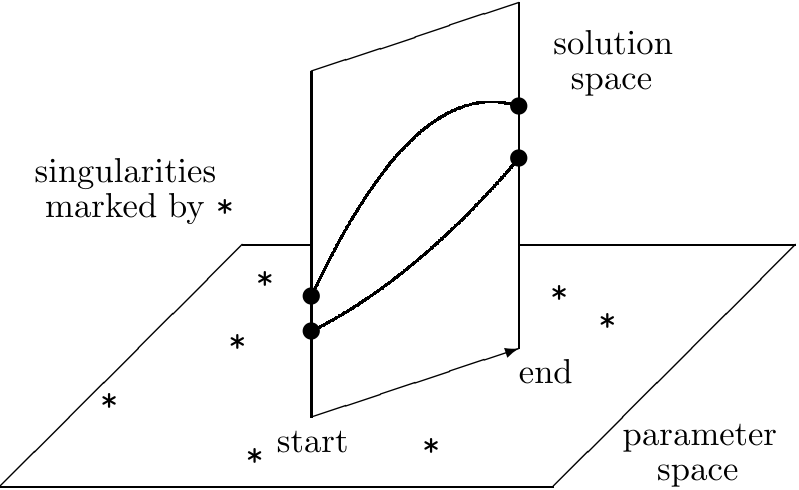}}
\caption{A generic choice for start parameters avoids singularities
along the paths.  \DUrole{label}{figcomplexparcon}}
\end{figure}

Solving a system that has solution sets of positive dimension, e.g.: a
curve or surface, means to compute its dimension and its degree. For a
solution set of dimension~$d$, we add to the system $d$
linear equations with random coefficients to reduce the problem to
computing isolated solutions.  If the dimension is not known in advance,
then it can be computed in a cascade of homotopies \cite{SV00}.
Because the coefficients of the linear
equation are random, the solutions of the system that satisfy the random
linear equations will be isolated. If the solution set has multiplicity
one, the solutions of the augmented system will be isolated points of
multiplicity one. Moreover, the number of isolated solutions of the
augmented system that lie on the $d$-dimensional solution set of
the original system will be equal to the degree of the
$d$-dimensional solution set. Thus a positive dimensional solution
set of dimension $d$ is represented by a set of $d$ random
linear equations and as many points in the intersection of the original
system with those random linear equations as the degree of the
$d$-dimensional solution set. In numerical algebraic geometry,
this representation is called a witness set.

For sparse polynomial systems with very few monomials appearing with
nonzero coefficient (in an extreme case, we consider binomial systems
that have exactly two monomials with nonzero coefficient in each
equation), we can represent positive dimensional solution sets by
monomial maps. For example, the two equations $x^2 y - zx = 0$,
$x^2 z - y^2 x = 0$ have as solutions three monomial maps:
$(x = 0, y = \lambda_1, z = \lambda_2)$,
$(x = \lambda_1, y = \lambda_1^2, z = \lambda_1^3)$, and
$(x = \lambda_1, y = 0, z = 0)$, for parameters $\lambda_1$
and $\lambda_2$. These monomial maps form the leading terms of
Puiseux series developments for general algebraic sets.

Surveys on homotopy continuation are \cite{AG93}, \cite{AG97}, \cite{Li03},
and \cite{Wat86}, \cite{Wat89}, \cite{Wat02}.
Book treatments are in \cite{AG03}, \cite{Mor87}, and \cite{SW05}.

\section{Interfaces to PHCpack and phc%
  \label{interfaces-to-phcpack-and-phc}%
}

This paper is mainly concerned with software problems.
There are at least three motivations to develop phcpy:\newcounter{listcnt0}
\begin{list}{\arabic{listcnt0}.}
{
\usecounter{listcnt0}
\setlength{\rightmargin}{\leftmargin}
}

\item 

PHCpack is a large Ada package, its executable phc
operates via menus, with input and output to files.
With phcpy we provide an interpreter interface to phc.
\item 

The code in PHCpack lacks adequate \emph{user} documentation
so that many of its features are not obviously accessible to users.
The Python modules of phcpy refactor the functionality of PHCpack
and beautiful documentation is generated by Sphinx \cite{Bra}.
\item 

As many new algorithms were first implemented with PHCpack,
reproducibility \cite{SBB13} of published computational results
can be automated via regression tests with Python scripts.\end{list}

Because also other interfaces to PHCpack may accomplish the same goals
outlined above, we first give an overview of the interfaces to PHCpack.

The first interface to PHCpack was based on the OpenXM~\cite{MNO+11} protocol
for the interaction of software components.
The virtue of this protocol is that only an executable version of the
software is required and one does not need to compile the code.

The interfaces to PHCpack from Maple~\cite{LV04}, MATLAB \& Octave~\cite{GV08b},
and Macaulay2~\cite{GPV13}
only require the executable~phc. This type of interface works in three
stages: (1) prepare an input file for phc; (2) call phc with some
options, the input file, and the name of an output file; (3) parse the
output file to extract the results. In principle, everything that can be
done via the command-line menus of phc can thus also be performed via
Maple procedures, MATLAB, Octave, or Macaulay2 scripts.

Figure~\DUrole{ref}{fighoney} shows the interfaces to PHCpack.\begin{figure}[]\noindent\makebox[\columnwidth][c]{\includegraphics[width=\columnwidth]{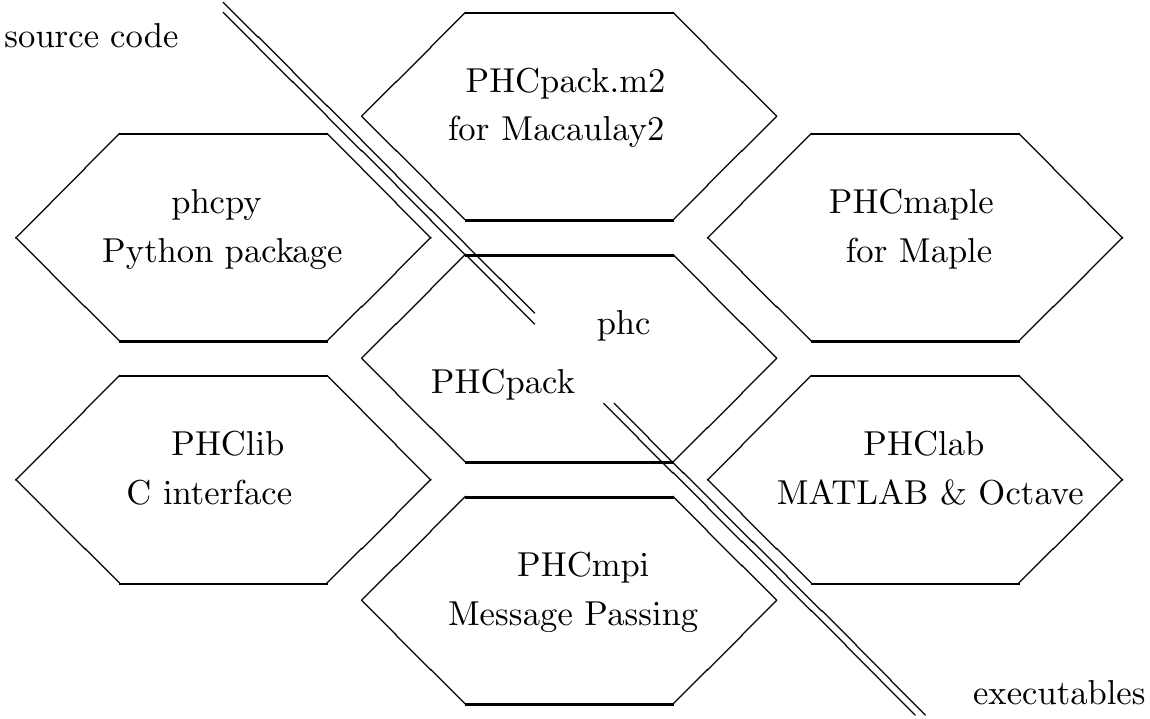}}
\caption{Diagram of the interfaces to PHCpack and phc.
The interfaces PHCpack.m2, PHCmaple, PHClab, depicted to the right of the
antidiagonal line require only the executable version phc.
The other interfaces PHClib, PHCmpi, and phcpy are based on the source
code of PHCpack.  \DUrole{label}{fighoney}}
\end{figure}

The C interface to PHCpack, described in~\cite{LV06}, offers the C programmer
access to the path trackers of PHCpack. This interface was developed for
use with the Message Passing Interface~\cite{SOH+98} and serves also as the basis
for~phcpy.  In the C interface, the data structures for polynomials and
solutions are not duplicated.  Instead of data structure duplication, one
can enter into the C interface routine a polynomial term after term.
The interface then behaves like a state machine.

Why would phcpy~be any better than the other interfaces? Leaving aside
the growing popularity of Python for scientific computing, the
replacement of files by persistent objects enabled the implementation of
a generator for the path trackers. After initialization of the homotopy
(with target, start system, and one start solution), the user can call a
“next” function to compute the next point at the solution path that
originates at the start solution given at initialization. This “next”
function (available for standard double, double double, quad double
precision, and arbitrary multiprecision) allows a detailed investigation
of the properties of a particular solution path.
In addition, it gives the user a fine control over the order of execution.
If desired, the tolerances and the step size can be adjusted as needed in
an application that plots solution trajectories.

Another (future) application of phcpy~is a web interface, such as at
\url{https://kepler.math.uic.edu} (beta version) presented by Xiangcheng Yu at
the SIAM AG 2013 conference in the first week of August 2013.

\section{Using phcpy%
  \label{using-phcpy}%
}

The blackbox solver of PHCpack is its most widely used function. In
phcpy, this blackbox solver is available in the function solve of the
module solver. The solver takes on input a list of strings that contain
valid representations of polynomials. On return is a list of strings,
which contain the solutions of the system.\begin{Verbatim}[commandchars=\\\{\},fontsize=\footnotesize]
\PY{o}{\PYZgt{}\PYZgt{}}\PY{o}{\PYZgt{}} \PY{k+kn}{from} \PY{n+nn}{phcpy.solver} \PY{k+kn}{import} \PY{n}{solve}
\PY{o}{\PYZgt{}\PYZgt{}}\PY{o}{\PYZgt{}} \PY{k+kn}{from} \PY{n+nn}{phcpy.phcpy2c} \PY{k+kn}{import} \PY{n}{py2c\PYZus{}set\PYZus{}seed}
\PY{o}{\PYZgt{}\PYZgt{}}\PY{o}{\PYZgt{}} \PY{n}{f} \PY{o}{=} \PY{p}{[}\PY{l+s}{\PYZdq{}}\PY{l+s}{x**2*y**2 + x + y;}\PY{l+s}{\PYZdq{}}\PY{p}{,}\PY{l+s}{\PYZdq{}}\PY{l+s}{x*y + x + y + 1;}\PY{l+s}{\PYZdq{}}\PY{p}{]}
\PY{o}{\PYZgt{}\PYZgt{}}\PY{o}{\PYZgt{}} \PY{n}{py2c\PYZus{}set\PYZus{}seed}\PY{p}{(}\PY{l+m+mi}{21320}\PY{p}{)}
\PY{l+m+mi}{0}
\PY{o}{\PYZgt{}\PYZgt{}}\PY{o}{\PYZgt{}} \PY{n}{s} \PY{o}{=} \PY{n}{solve}\PY{p}{(}\PY{n}{f}\PY{p}{,}\PY{n}{silent}\PY{o}{=}\PY{n+nb+bp}{True}\PY{p}{)}
\PY{o}{\PYZgt{}\PYZgt{}}\PY{o}{\PYZgt{}} \PY{n+nb}{len}\PY{p}{(}\PY{n}{s}\PY{p}{)}
\PY{l+m+mi}{4}
\PY{o}{\PYZgt{}\PYZgt{}}\PY{o}{\PYZgt{}} \PY{k}{print} \PY{n}{s}\PY{p}{[}\PY{l+m+mi}{0}\PY{p}{]}
\PY{n}{t} \PY{p}{:} \PY{l+m+mf}{1.00000000000000E+00} \PY{l+m+mf}{0.00000000000000E+00}
\PY{n}{m} \PY{p}{:} \PY{l+m+mi}{1}
\PY{n}{the} \PY{n}{solution} \PY{k}{for} \PY{n}{t} \PY{p}{:}
\PY{n}{x} \PY{p}{:} \PY{o}{\PYZhy{}}\PY{l+m+mf}{1.00000000000000E+00} \PY{l+m+mf}{0.00000000000000E+00}
\PY{n}{y} \PY{p}{:} \PY{o}{\PYZhy{}}\PY{l+m+mf}{1.61803398874989E+00} \PY{l+m+mf}{0.00000000000000E+00}
\PY{o}{==} \PY{n}{err} \PY{p}{:} \PY{l+m+mf}{2.143E\PYZhy{}101} \PY{o}{=} \PY{n}{rco} \PY{p}{:} \PY{l+m+mf}{4.775E\PYZhy{}02} \PY{o}{=} \PY{n}{res} \PY{p}{:} \PY{l+m+mf}{2.220E\PYZhy{}16} \PY{o}{=}
\end{Verbatim}
With py2c\_set\_seed() we fix the seed of the random number generator
for the coefficients of the start system in the homotopy, which makes
for predictable runs.  Otherwise, the solve() each time generates
different coefficients in the homotopies and the order of the solutions
on return may differ.
For each solution, the triplet (err,rco,res) indicates the quality of
the solution:%
\begin{itemize}

\item 

err: the norm of the last update made by Newton’s method (forward
error),
\item 

rco: estimate for the inverse condition number of the Jacobian
matrix,
\item 

res: norm of the evaluated solution (backward error).
\end{itemize}

With double double and quad double arithmetic we get more accurate
solutions.

To predict the number of isolated solutions with the mixed volume:\begin{Verbatim}[commandchars=\\\{\},fontsize=\footnotesize]
\PY{o}{\PYZgt{}\PYZgt{}}\PY{o}{\PYZgt{}} \PY{k+kn}{from} \PY{n+nn}{phcpy.solver} \PY{k+kn}{import} \PY{n}{mixed\PYZus{}volume}
\PY{o}{\PYZgt{}\PYZgt{}}\PY{o}{\PYZgt{}} \PY{n}{mixed\PYZus{}volume}\PY{p}{(}\PY{n}{f}\PY{p}{)}
\PY{l+m+mi}{4}
\end{Verbatim}
Version 0.1.5 of phcpy~contains the following modules:%
\begin{itemize}

\item 

solver: a blackbox solver, mixed-volume calculator, linear-product
root count and start system, path trackers, deflation for isolated
singular solutions.
\item 

examples: a selection of interesting benchmark systems.
Typing python examples.py at the command prompt calls the
blackbox solver on all benchmark examples, thus providing
an automatic regression test.
\item 

families: some problems can be formulated for any number of
variables.
\item 

phcmaps: monomial maps as solutions of binomial systems.
\item 

phcsols: conversion of PHCpack solution strings into Python
dictionaries.
\item 

phcsets: basic tools to manipulate positive dimensional solution
sets.
\item 

phcwulf: basic client/server setup to solve many systems.
\item 

schubert: the Pieri homotopies solve particular polynomial systems
arising in enumerative geometry.
\end{itemize}

The number of exported functions, documented by Sphinx~\cite{Bra} runs in the
several hundreds. The code of version 0.1.1 of phcpy was improved with
the aid of Pylint~\cite{The}, yielding a global rating of~9.73/10.

\section{The Design of phcpy%
  \label{the-design-of-phcpy}%
}

The design of phcpy~is drawn in Figure~\DUrole{ref}{figphcpy}. This design can be
viewed as an application of a façade pattern~(see Figure B.31 in \cite{Bai08}).
The façade pattern plays
a strong role in converting legacy systems incrementally to more modern
software and is appropriate as phcpy~should be viewed as a modernization
of PHCpack. The implementation of use\_c2phc.adb applies the chain of
responsibility pattern~(see Figure B.12 in \cite{Bai08}),
calling handlers to specific packages in
PHCpack. That we use the name phcpy~and not PyPHC indicates that phcpy
is more than just an interface.\begin{figure}[]\noindent\makebox[\columnwidth][c]{\includegraphics[width=\columnwidth]{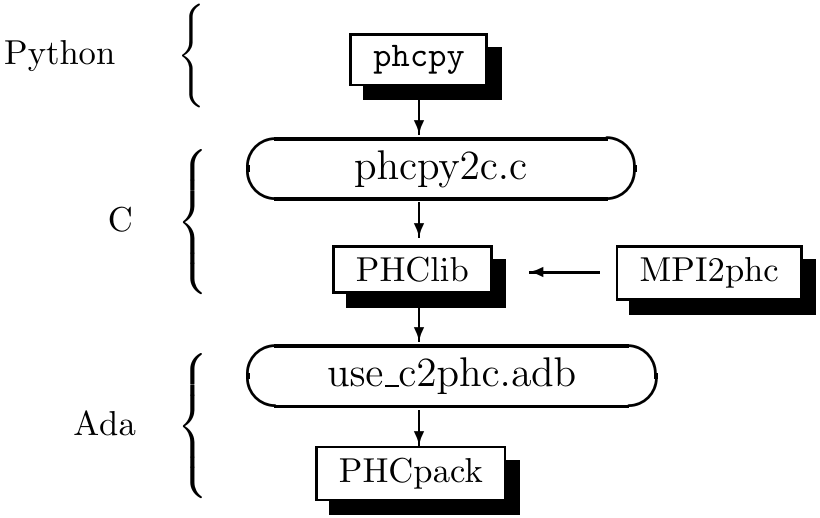}}
\caption{The design of phcpy depends on PHClib, a library of various
collections of C functions, through one file phcpy2c.c
(with documentation in the corresponding header phcpy2c.h)
which encodes the Python bindings.
PHClib interfaces to the Ada routines of PHCpack
through one Ada procedure use\_c2phc.adb.
The collection of parallel programs (MPI2phc)
using message passing (MPI) depends on PHClib.  \DUrole{label}{figphcpy}}
\end{figure}

The code for phcpy~builds directly on the C interface to PHCpack.
The C interface was developed to use the Message Passing Interface
(MPI)~\cite{SOH+98}.
In joint work with Yusong Wang~\cite{VW02}, \cite{VW04a}, \cite{VW04b},
Yan Zhuang~\cite{VZ06}, Yun Guan~\cite{GV08a},
and Anton Leykin \cite{LV05}, \cite{LV09}, \cite{LVZ06},
the main program was always a C program.
The C interface described in~\cite{LV06}
is centered around one gateway function use\_c2phc.
To the Ada programmer, this function has the specification\begin{Verbatim}[commandchars=\\\{\},fontsize=\footnotesize]
\PY{k+kd}{function} \PY{n+nf}{use\PYZus{}c2phc} \PY{p}{(} \PY{n+nv}{job} \PY{p}{: }\PY{n+nv}{integer}\PY{p}{;}
                     \PY{n+nv}{a} \PY{p}{: }\PY{n+nv}{C\PYZus{}intarrs}\PY{p}{.}\PY{n+nv}{Pointer}\PY{p}{;}
                     \PY{n+nv}{b} \PY{p}{: }\PY{n+nv}{C\PYZus{}intarrs}\PY{p}{.}\PY{n+nv}{Pointer}\PY{p}{;}
                     \PY{n+nv}{c} \PY{p}{: }\PY{n+nv}{C\PYZus{}dblarrs}\PY{p}{.}\PY{n+nv}{Pointer} \PY{p}{)}
                   \PY{k+kr}{return} \PY{k+kt}{integer}\PY{p}{;}
\end{Verbatim}
The prototype of the corresponding C function is\begin{Verbatim}[commandchars=\\\{\},fontsize=\footnotesize]
\PY{k}{extern} \PY{k+kt}{int} \PY{n+nf}{\PYZus{}ada\PYZus{}use\PYZus{}c2phc} \PY{p}{(} \PY{k+kt}{int} \PY{n}{task}\PY{p}{,}
                            \PY{k+kt}{int} \PY{o}{*}\PY{n}{a}\PY{p}{,}
                            \PY{k+kt}{int} \PY{o}{*}\PY{n}{b}\PY{p}{,}
                            \PY{k+kt}{double} \PY{o}{*}\PY{n}{c} \PY{p}{)}\PY{p}{;}
\end{Verbatim}
With use\_c2phc we obtain one uniform streamlined design of the
interface: the C programmer calls one single Ada function
\_ada\_use\_c2phc. What use\_c2phc executes depends on the job number.
The (a,b,c) parameters are flexible enough to pass strings
and still provide some form of type checking (which would not
be possible had we wiped out all types with void*).

To make \_ada\_use\_c2phc usable, we have written a number of C
wrappers, responsible for parsing the arguments of the C functions to be
passed to \_ada\_use\_c2phc. The extension module and the shared object
for the implementation of phcpy is a set of wrappers defined by
phcpy2c.c and documented by phcpy2c.h. As a deliberate design decision
of phcpy, all calls to functions in PHCpack pass through the C
interface. By this design, the development of phcpy benefits the C and
C++ programmers.

\section{Obtaining, Installing, and Contributing%
  \label{obtaining-installing-and-contributing}%
}

PHCpack and phcpy are distributed under the GNU GPL license
(version 2 or any later version).
Recently a new repository PHCpack was added on github
with the source code of version 2.3.85 of PHCpack,
which contains version 0.1.5 of phcpy.
Executable versions for Linux, Mac, and Windows are
available via the homepage of the author.

The code was developed on a Red Hat Enterprise Linux Workstation
(Release 6.4) and a MacBook Pro laptop (Mac OS X 10.8.5)
using the GNAT GPL 2013 compiler.
Versions 2.6.6 and 2.7.3 of Python, respectively on Linux and Mac,
were used to develop phcpy.  Packaged binary distributions of
phcpy for the platforms listed above are available via the
homepage of the author.

Although the blackbox solver of PHCpack has been in use since 1996,
phcpy itself is still very much in beta stage.
Suggestions for improvement and contributions to phcpy
will be greatly appreciated.

\section{Acknowledgments%
  \label{acknowledgments}%
}

The author thanks Max Demenkov for his comments and questions
at the poster session at EuroSciPy 2013.  In particular the question
on obtaining all solutions along a path led to the introduction of
generator functions for the path trackers in version 0.1.4 of phcpy.

This material is based upon work supported by the National Science
Foundation under Grant No.~1115777.

\end{document}